\documentstyle{article}
\begin{document}\bibliographystyle{plain}
\begin{titlepage}\renewcommand{\thefootnote}{\fnsymbol{footnote}}
\large\hfill\begin{tabular}{l}HEPHY-PUB
725/99\\UWThPh-1999-77\\hep-ph/0001191\\December 1999\end{tabular}\\[4cm]
\Large\begin{center}{\bf COMMENT ON ``CALCULATION OF THE\\QUARKONIUM SPECTRUM
AND $\mbox{\boldmath{$m$}}_{\bf b},$ $\mbox{\boldmath{$m$}}_{\bf c}$\\TO
ORDER $\mbox{\boldmath{$\alpha$}}_{\bf s}^4$''}\\\vspace{2cm}\Large{\bf
Wolfgang LUCHA\footnote[1]{\normalsize\ {\em E-mail\/}:
wolfgang.lucha@oeaw.ac.at}}\\[.5cm]\large Institut f\"ur
Hochenergiephysik,\\\"Osterreichische Akademie der
Wissenschaften,\\Nikolsdorfergasse 18, A-1050 Wien, Austria\\[2cm]\Large {\bf
Franz F. SCH\"OBERL\footnote[2]{\normalsize\ {\em E-mail\/}:
franz.schoeberl@univie.ac.at}}\\[.5cm]\large Institut f\"ur Theoretische
Physik,\\Universit\"at Wien,\\Boltzmanngasse 5, A-1090 Wien, Austria

\vfill{\large\bf Abstract}\end{center}\normalsize A recent determination of
the mass of the b quark, based exclusively on quantum chromodynamics~(by
avoiding strictly to introduce any phenomenological interaction potential of
nonperturbative origin), may be improved by allowing for a merely numerical
solution of the corresponding eigenvalue problem.

\vspace{3ex}

\renewcommand{\thefootnote}{\arabic{footnote}}\end{titlepage}

\normalsize Recently, Pineda and Yndur\'ain \cite{PinedaYndurain98} presented
a re-analysis of heavy quarkonia. Their investigation is based on the main
assumption that bound systems of heavy quarks may be reasonably described by
nonrelativistic kinematics and only the perturbative contribution to the
quark--antiquark interaction potential $V$ if all nonperturbative effects are
taken into account by some appropriate correction to~the energy. In order to
describe a system of a heavy quark and antiquark, both with constituent mass
$m$, forming a bound state with total spin $s=0$ or $s=1$, Pineda and
Yndur\'ain consider the Hamiltonian
\begin{eqnarray}H&=&2\,m-\frac{1}{m}\,\Delta-\frac{1}{4\,m^3}\,\Delta^2
+V_0^{\rm P}(r)+\frac{C_{\rm F}\,\alpha(\mu)}{m^2\,r}\,\Delta+\frac{C_{\rm
F}\,\alpha^2(\mu)}{4\,m\,r^2}\,(C_{\rm F}-2\,C_{\rm A})\nonumber\\[1ex]
&+&\frac{4\,\pi\,C_{\rm F}\,\alpha(\mu)}{3\,m^2}\,s\,(s+1)\,\delta^{(3)}({\bf
x})\ ,\quad r\equiv|{\bf x}|\ .\label{Eq:HPY}\end{eqnarray}The perturbative
contribution to the static quark--antiquark interaction potential, $V_0^{\rm
P}(r),$ is known~up to and including the two-loop level
\cite{StatPot2L}:\begin{eqnarray*}V_0^{\rm P}(r) &=&-C_{\rm
F}\,\frac{\alpha(\mu)}{r}\left\{1+\frac{\alpha(\mu)}{4\,\pi}
\left[\frac{5}{3}\,\beta_0-\frac{8}{3}\,C_{\rm A}+2\,\beta_0\,
[\ln(\mu\,r)+\gamma_{\rm E}]\right]\right.\\[1ex]
&+&\left(\frac{\alpha(\mu)}{4\,\pi}\right)^2\left[
\beta_0^2\left(4\,[\ln(\mu\,r)+\gamma_{\rm
E}]^2+\frac{\pi^2}{3}\right)+\left(
2\,\beta_1+\frac{20}{3}\,\beta_0^2-\frac{32}{3}\,\beta_0\,C_{\rm A}\right)
[\ln(\mu\,r)+\gamma_{\rm E}]\right.\\[1ex]&+&\left(\frac{4343}{162}+4\,\pi^2
-\frac{\pi^4}{4}+\frac{22}{3}\,\zeta(3)\right)C_{\rm
A}^2-\left(\frac{1798}{81} +\frac{56}{3}\,\zeta(3)\right)n_{\rm f}\,C_{\rm
A}\,T_{\rm F}\\[1ex]&-&
\left.\left.\left(\frac{55}{3}-16\,\zeta(3)\right)n_{\rm f}\,C_{\rm F}\,
T_{\rm F}+\frac{400}{81}\,n_{\rm f}^2\,T_{\rm F}^2\right]\right\}.
\end{eqnarray*}Here, the following notations have been adopted: $\alpha(\mu)$
denotes the strong fine-structure constant~in the modified
minimal-subtraction ($\overline{\rm MS}$) renormalization scheme. For a
non-Abelian gauge theory for $n_{\rm f}$ Dirac fermions, invariant
w.~r.~t.~gauge transformations forming a Lie group SU($N$) describing $N$
colour~degrees of freedom, the quadratic Casimir invariants read, for the
fundamental representation,$$C_{\rm F}=\frac{N^2-1}{2\,N}$$and, for the
adjoint representation,$$C_{\rm A}=N\ ,$$if the generators of the Lie group
SU($N$) are normalized such that the second-order Dynkin index~of the
fundamental representation is$$T_{\rm F}=\frac{1}{2}\ .$$The dependence of
the effective (running) fine-structure constant $\alpha(\mu)$ on the
renormalization scale~$\mu$ is described in terms of the Gell-Mann--Low
$\beta$ function according
to$$\frac{\mu}{2}\,\frac{\partial}{\partial\mu}\alpha(\mu)=
-\frac{\alpha^2(\mu)}{4\,\pi}\,\beta_0
-\frac{\alpha^3(\mu)}{(4\,\pi)^2}\,\beta_1
-\frac{\alpha^4(\mu)}{(4\,\pi)^3}\,\beta_2+O(\alpha^5)\ ,$$involving the
well-known expressions for the (gauge-invariant) one-, two-, and three-loop
expansion coefficients in the $\overline{\rm MS}$ scheme
\cite{Tarasov80}\begin{eqnarray*}
\beta_0&=&\frac{11}{3}\,C_{\rm A}-\frac{4}{3}\,n_{\rm f}\,T_{\rm F}\ ,\\[1ex]
\beta_1&=&\frac{34}{3}\,C_{\rm A}^2-\frac{20}{3}\,n_{\rm f}\,C_{\rm
A}\,T_{\rm F}-4\,n_{\rm f}\,C_{\rm F}\,T_{\rm F}\ ,\\[1ex]
\beta_2&=&\frac{2857}{54}\,C_{\rm A}^3-\frac{1415}{27}\,n_{\rm f}\,C_{\rm
A}^2\,T_{\rm F}+\frac{158}{27}\,n_{\rm f}^2\,C_{\rm A}\,T_{\rm F}^2
-\frac{205}{9}\,n_{\rm f}\,C_{\rm A}\,C_{\rm F}\,T_{\rm
F}+\frac{44}{9}\,n_{\rm f}^2\,C_{\rm F}\,T_{\rm F}^2 +2\,n_{\rm f}\,C_{\rm
F}^2\,T_{\rm F}\ .\end{eqnarray*}The resulting dependence of the
fine-structure constant $\alpha(\mu)$ on the chosen renormalization scale
$\mu$, expressed in terms of the (standard) scale parameter $\Lambda$, reads
up to and including the three-loop~level\begin{eqnarray*}\alpha(\mu)&=&
\frac{4\,\pi}{\beta_0\ln(\mu^2/\Lambda^2)}\left\{1-\frac{\beta_1}{\beta_0^2}\,
\frac{\ln(\ln(\mu^2/\Lambda^2))}{\ln(\mu^2/\Lambda^2)}\right.\\[1ex]
&+&\left.\frac{\beta_1^2}{\beta_0^4\ln^2(\mu^2/\Lambda^2)}
\left[\left(\ln(\ln(\mu^2/\Lambda^2))-\frac{1}{2}\right)^2
+\frac{\beta_2\,\beta_0}{\beta_1^2}-\frac{5}{4}\right]\right\}.
\end{eqnarray*}$\gamma_{\rm E}$ is known as Euler--Mascheroni constant. In
the case of quantum chromodynamics, clearly, $N=3.$

\newpage

Now, in order to stick to an entirely analytical analysis and following the
philosophy developed~in an earlier treatment \cite{TitardYndurain94} of heavy
quarkonia, in Ref.~\cite{PinedaYndurain98} the static potential $V_0^{\rm
P}(r)$ is~split, according~to\begin{equation}V_0^{\rm P}(r)=\tilde V(r)+\hat
V(r)\ ,\label{Eq:statpotsplit}\end{equation}into the Coulomb-like
contribution$$\tilde V(r)\equiv-C_{\rm F}\,\frac{\tilde\alpha(\mu)}{r}\
,$$with the effective fine-structure
constant\begin{eqnarray*}\tilde\alpha(\mu)&\equiv&\alpha(\mu)
\left\{1+\frac{\alpha(\mu)}{4\,\pi}\left(\frac{5}{3}\,\beta_0
-\frac{8}{3}\,C_{\rm A}+2\,\beta_0\,\gamma_{\rm E}\right)\right.\\[1ex]
&+&\left(\frac{\alpha(\mu)}{4\,\pi}\right)^2\left[
\beta_0^2\left(4\,\gamma_{\rm E}^2+\frac{\pi^2}{3}\right)+\left(
2\,\beta_1+\frac{20}{3}\,\beta_0^2-\frac{32}{3}\,\beta_0\,C_{\rm A}\right)
\gamma_{\rm E}\right.\\[1ex]&+&\left(\frac{4343}{162}+4\,\pi^2
-\frac{\pi^4}{4}+\frac{22}{3}\,\zeta(3)\right)C_{\rm
A}^2-\left(\frac{1798}{81} +\frac{56}{3}\,\zeta(3)\right)n_{\rm f}\,C_{\rm
A}\,T_{\rm F}\\[1ex]&-&
\left.\left.\left(\frac{55}{3}-16\,\zeta(3)\right)n_{\rm f}\,C_{\rm F}\,
T_{\rm F}+\frac{400}{81}\,n_{\rm f}^2\,T_{\rm F}^2\right]\right\},
\end{eqnarray*}and an obvious remainder involving logarithms of the radial
coordinate $r$,\begin{eqnarray*}\hat V(r)&=&-C_{\rm
F}\,\frac{\alpha^2(\mu)}{4\,\pi\,r}
\left\{2\,\beta_0\,\ln(\mu\,r)\frac{}{}\right.\\[1ex]
&+&\left.\frac{\alpha(\mu)}{4\,\pi}\left[4\,\beta_0^2\,\ln^2(\mu\,r)
+\left(2\,\beta_1+\frac{20}{3}\,\beta_0^2-\frac{32}{3}\,\beta_0\,C_{\rm A}
+8\,\beta_0^2\,\gamma_{\rm
E}\right)\ln(\mu\,r)\right]\right\}.\end{eqnarray*}The eigenvalue problem for
the Coulombic Hamiltonian $$\tilde H\equiv 2\,m-\frac{1}{m}\,\Delta+\tilde
V(r)$$is solved exactly, yielding, for instance, for the ground state, the
energy eigenvalue $\tilde E_0=2\,m+\varepsilon_0,$ with the Coulomb binding
energy$$\varepsilon_0=-\frac{C_{\rm F}^2\,\tilde\alpha^2(\mu)\,m}{4}\ .$$The
``non-Coulombic'' part $H-\tilde H$ of the Hamiltonian (\ref{Eq:HPY}) is
treated perturbatively. Counting carefully the powers of $\alpha$ yields
analytical expressions for the eigenvalues of $H$ correct up to and including
the order $\alpha^4$. Nonperturbative effects are incorporated by adding the
Leutwyler--Voloshin correction \cite{NPC}, which involves the gluon
condensate $\langle\alpha\,G^2\rangle=0.06\pm 0.02\;\mbox{GeV}^4$. For the
ground state, this correction amounts to the energy shift$$\delta
E_0=\frac{624}{425}\,\frac{\pi\,\langle\alpha\,G^2\rangle\,m}{(C_{\rm
F}\,\tilde\alpha(\mu)\,m)^4}\ .$$By inversion of the expression for the
(ground-state) bound-state mass emerging from this procedure, the
corresponding quark pole mass $m$ is computed. For instance, with
$\Lambda(n_{\rm
f}=4)=0.23\!\!{\tiny\begin{array}{l}+0.08\\-0.05\end{array}}\!\mbox{GeV}$~and
$\mu=\sqrt{6.632\pm 25\,\%}\;\mbox{GeV},$ implying $\alpha(\mu)=0.246$ and
$\tilde\alpha(\mu)=0.386,$ the experimental value \cite{PDT98} of the
$\Upsilon$ mass, $M(\Upsilon)_{\rm exp}=9.46037\pm 0.00021\;\mbox{GeV},$
translates into the b quark mass $m_{\rm b}=5.001\!\!{\tiny\begin{array}{l}
+0.104\\-0.066\end{array}}\!\mbox{GeV}$~\cite{PinedaYndurain98}.

However, a perturbative treatment as implied by the splitting
(\ref{Eq:statpotsplit}) of the static potential $V_0^{\rm P}(r)$~is by no
means mandatory, obligatory, or even desirable. We may also adopt the
following point of~view. Given the operator $H$ defined by Eq.~(\ref{Eq:HPY})
(accurate up to a certain order in $\alpha$), compute (numerically,~if
necessary) its discrete spectrum, i.~e., the set of eigenvalues, irrespective
of the involved powers of~$\alpha$. Of course, the terms in
Eq.~(\ref{Eq:HPY}) proportional to$$-\Delta^2\ ,\quad+\frac{1}{r}\,\Delta\
,$$and, if the effective coupling strength multiplying this term exceeds some
critical value, also the term in Eq.~(\ref{Eq:HPY}) proportional
to$$-\frac{1}{r^2}$$render the operator $H$ unbounded from below and have
therefore to be treated perturbatively anyway. The
Hamiltonian\begin{equation}\hat H\equiv 2\,m-\frac{1}{m}\,\Delta+V_0^{\rm
P}(r)\ ,\label{Eq:SPH}\end{equation}on the other hand, may certainly be
analyzed without adhering to some perturbative approximation.

\newpage

In order to obtain a first idea of the differences brought about by these two
approaches, let us~start by considering only the Hamiltonian (\ref{Eq:SPH}).
The perturbative calculation is straightforward.~Introducing, for notational
brevity, the generalized Bohr radius $$a(\mu)\equiv\frac{2}{C_{\rm
F}\,\tilde\alpha(\mu)\,m}\ ,$$the expectation values of the non-Coulombic
interaction $\hat V(r)$ w.~r.~t.~the ground state of $\tilde H$ (indicated by
the subscript $\tilde 0$) may be evaluated with the help of the relations
(see also Appendix~B of Ref.~\cite{TitardYndurain94})
\begin{eqnarray*}\left\langle\frac{\ln(\mu\,r)}{r}\right\rangle_{\tilde 0}&=&
\frac{1}{a(\mu)}\left[\ln\left(\frac{\mu\,a(\mu)}{2}\right)+1-\gamma_{\rm
E}\right],\\[1ex]\left\langle\frac{\ln^2(\mu\,r)}{r}\right\rangle_{\tilde 0}&=&
\frac{1}{a(\mu)}\left[\ln^2\left(\frac{\mu\,a(\mu)}{2}\right)+2\,(1-\gamma_{\rm
E})\,\ln\left(\frac{\mu\,a(\mu)}{2}\right)+(1-\gamma_{\rm
E})^2+\frac{\pi^2}{6}-1\right].\end{eqnarray*}The nonperturbative evaluation
of $\hat H$ is performed with some numerical procedure\footnote{\ The desired
accuracy of the (numerically determined) bound-state energies and wave
functions may be adjusted~in the routine used for the solution of the
Schr\"odinger equation. For the present analysis, the uncertainty of these
energies has been required to be less than
$10^{-7}\;\mbox{GeV}=100\;\mbox{eV}.$ This accuracy should be, by far,
sufficient for our purposes.} developed for~the treatment of the
nonrelativistic Schr\"odinger equation \cite{Lucha98IJMPC}. In this way, we
find, for the parameter values used in the second of
Refs.~\cite{PinedaYndurain98} and focusing our interest to the ground state,
for the Coulomb binding energy, $\varepsilon_0=-0.33146\;\mbox{GeV},$ for the
expectation value of $\hat V(r)$, $\langle\hat V(r)\rangle_{\tilde
0}=-0.13714\;\mbox{GeV},$ and thus, for the perturbatively calculated
ground-state energy,\footnote{\ For the perturbative treatment of $\hat H$,
we truncate the Rayleigh--Schr\"odinger series for $\hat E_0^{\rm P}$ at
lowest non-trivial~order in $\hat V(r)$. Inclusion of the next order
\cite{PinedaYndurain98} reduces the observed discrepancies but does not
change qualitatively our findings.} $\hat E_0^{\rm P}\equiv
2\,m+\varepsilon_0+\langle\hat V(r)\rangle_{\tilde 0}=9.5334\;\mbox{GeV},$
while the numerically computed ``exact'' lowest eigenvalue of the operator
(\ref{Eq:SPH}) is $\hat E_0^{\rm NP}\equiv\langle\hat
H\rangle_0=9.5198\;\mbox{GeV}.$ Consequently, for the parameters of
Ref.~\cite{PinedaYndurain98} the difference in the lowest bound-state mass
predicted by the Hamiltonian (\ref{Eq:SPH}) within perturbative and
nonperturbative approaches is $\hat E_0^{\rm P}-\hat E_0^{\rm
NP}=13.6\;\mbox{MeV}.$ Unfortunately, this discrepancy is roughly 65 times
larger than the experimental error on the $\Upsilon$ mass. Phrased the other
way round, the perturbative ground-state eigenvalue $\hat E_0^{\rm P}$ of
$\hat H$ can be reproduced~by the nonperturbative evaluation of $\hat H$ for
a mass $m$ of the bound-state constituents of $m=5.008\;\mbox{GeV}.$ For the
b quark mass of Ref.~\cite{PinedaYndurain98}, the theoretical error different
from the one induced by variation of the renormalization scale $\mu,$
attributed to neglected higher-order perturbative as well as nonperturbative
corrections, is estimated to be $\pm0.006\;\mbox{GeV}.$ Obviously, this error
is entirely consumed already by~the difference of the masses of the
bound-state constituents obtained by perturbative and nonperturbative
evaluations of $\hat H.$

In view of these findings, let's try to improve the theoretical value of the
b quark pole mass $m_{\rm b}$~by approaching the part $\hat H$ of the
Hamiltonian (\ref{Eq:HPY}) nonperturbatively. The numerical computation
of~the expectation values of the operator $H-\hat H$ is considerably
facilitated by the following two observations:\begin{itemize}\item The
eigenvalue equation for the ``toy-model'' (Hamiltonian) operator $\hat H$
defined in Eq.~(\ref{Eq:SPH}) reads, for some generic (energy) eigenvalue
$\hat E$ and its corresponding eigenstate $|\psi\rangle$ of $\hat H$, $\hat
H|\psi\rangle=\hat E\,|\psi\rangle.$ Thus the expectation values of all terms
in the Hamiltonian (\ref{Eq:HPY}) involving the Laplacian $\Delta,$ taken
w.~r.~t.~$|\psi\rangle,$ may be evaluated by substituting $\Delta$ according
to $\Delta|\psi\rangle=m\left[2\,m+V_0^{\rm P}(r)-\hat
E\right]|\psi\rangle.$\item The expectation value of the $\delta$ function
entering in the ``spin--spin term'' of the Hamiltonian~(\ref{Eq:HPY}), taken
w.~r.~t.~$|\psi\rangle,$ is the modulus squared of the corresponding wave
function $\psi({\bf x})$ at the origin: $\langle\psi|\delta^{(3)}({\bf
x})|\psi\rangle=|\psi({\bf 0})|^2.$ For states with vanishing orbital angular
momentum $\ell$ (the so-called ``S waves''), $|\psi({\bf 0})|^2$ may be
expressed in terms of the first derivative of the relevant interaction
potential $V(r)$ w.~r.~t.~the radial coordinate $r$ according to (for a
derivation, see, e.~g., Ref.~\cite{Lucha91PRep})$$|\psi({\bf
0})|^2=\frac{m}{4\,\pi}
\left\langle\psi\left|\frac{{\rm d}}{{\rm d}r}V(r)\right|\psi\right\rangle.$$
\end{itemize}In this way, the lowest eigenvalue of the Hamiltonian $H$ may be
computed from the expression (where the subscript 0 of the expectation values
indicates, as before, the ground state of the Hamiltonian~$\hat H$)
\begin{eqnarray*}E_0\equiv\langle H\rangle_0&=&\hat E_0^{\rm
NP}-\frac{1}{4\,m^3}\left\langle\Delta^2\right\rangle_0+\frac{C_{\rm
F}\,\alpha(\mu)}{m^2}\left\langle\frac{1}{r}\,\Delta\right\rangle_0
+\frac{C_{\rm F}\,\alpha^2(\mu)}{4\,m}\,(C_{\rm F}-2\,C_{\rm
A})\left\langle\frac{1}{r^2}\right\rangle_0\\[1ex] &+&\frac{4\,\pi\,C_{\rm
F}\,\alpha(\mu)}{3\,m^2}\,s\,(s+1)\,|\psi_0({\bf 0})|^2\
.\end{eqnarray*}Adding the nonperturbative shift $\delta E_0$ gives our final
result for the ground-state energy: ${\cal E}_0=E_0+\delta E_0.$ For the
parameter values of Ref.~\cite{PinedaYndurain98} and a b quark mass of
$m_{\rm b}=5.001\;\mbox{GeV},$ this expression entails~the bound-state energy
${\cal E}_0=9.4953\;\mbox{GeV}.$ Hence, the error of the predicted $\Upsilon$
mass brought about by~the perturbative approximation in the treatment of the
Hamiltonian (\ref{Eq:HPY}), which amounts to $35\;\mbox{MeV},$ is~of the
order of the hyperfine splittings in the bottomonium system. For the latter,
the second of Refs.~\cite{PinedaYndurain98} quotes, e.~g.,
$M(\Upsilon)-M(\eta_{\rm
b})=46.6\!\!{\tiny\begin{array}{l}+14.8\\-12.7\end{array}}\!\mbox{MeV}.$

Consequently, a re-evaluation of the b quark mass appears to be in order.
Fitting the ground-state bound-state energy ${\cal E}_0$, for the numerical
values of the parameters $\Lambda,$ $\mu,$ and $\langle\alpha\,G^2\rangle$
adopted~in~Ref.~\cite{PinedaYndurain98}, to the experimental mass of the
$\Upsilon,$ we obtain for the pole mass of the b quark$$m_{\rm
b}=4.983\;\mbox{GeV}\ .$$The errors caused by the uncertainties of $\Lambda$
and $\langle\alpha\,G^2\rangle$ and by the variation of $\mu$ should be
practically the same as in Ref.~\cite{PinedaYndurain98}. Hence, dropping the
requirement of analytical accessibility of the Hamiltonian (\ref{Eq:HPY})
reduces the extracted b quark mass by some $18\;\mbox{MeV}.$ A very similar
result is expected to be~found for the determination of the c quark mass. We
arrive at the conclusion that in Refs.~\cite{PinedaYndurain98} the
theoretical errors on the quark masses have been somewhat underestimated.

Note that the above considerations apply, of course, also to the analysis
presented in Ref.~\cite{TitardYndurain94}.~Note also that the correct
\cite{Tarasov80} numerical factor in the numerator of the first term in the
expression for the three-loop $\beta$ function coefficient $\beta_2$ differs
slightly from the one used in Refs.~\cite{PinedaYndurain98}.

\end{document}